\documentclass{emulateapj}

\def\gtsima{$\; \buildrel > \over \sim \;$}
\def\ltsima{$\; \buildrel < \over \sim \;$}
\def\prosima{$\; \buildrel \propto \over \sim \;$}
\def\gsim{\lower.5ex\hbox{\gtsima}}
\def\lsim{\lower.5ex\hbox{\ltsima}}
\def\simgt{\lower.5ex\hbox{\gtsima}}
\def\simlt{\lower.5ex\hbox{\ltsima}}
\def\simpr{\lower.5ex\hbox{\prosima}}

\def\h1{$h^{-1}$}
\def\eeq{\end{equation}}
\def\beq{\begin{equation}}

\shorttitle{Near-IR luminous galaxies at z$\sim$2}
\shortauthors{de Mello et al.}

\begin{document}


\title{Metal enrichment in near-IR luminous galaxies at \lowercase{$z\sim 2$}:
signatures of proto-ellipticals?\footnote{Based on observations taken with the
European Southern Observatory, Chile, ESO programs 70.A-0140 and 168.A-0485}}

\author{D.F. de Mello\altaffilmark{2,3}, E. Daddi\altaffilmark{4}, 
A. Renzini\altaffilmark{4}, 
A. Cimatti\altaffilmark{5},
S. di Serego Alighieri\altaffilmark{5},
L. Pozzetti\altaffilmark{6}, 
G. Zamorani\altaffilmark{6}, 
} 

\altaffiltext{2}{LASP, Code 681, GSFC, Greenbelt, MD 20771, USA}
\altaffiltext{3}{Catholic University of America, Washington, DC 20064, USA}
\altaffiltext{4}{ESO, Karl-Schwarzschild-Str. 2, D-85748 Garching, Germany}
\altaffiltext{5}{INAF-Oss. Astr. di Arcetri, L.go E.Fermi 5, Firenze, Italy}
\altaffiltext{6}{INAF-Oss. Astr. di Bologna, via Ranzani 1, Bologna, Italy}




\begin{abstract}
We present the analysis of the coadded rest-frame UV spectrum ($1200<z<2000$ \AA)
of five $K$-luminous galaxies at $z\sim 2$ from the K20 survey. The
composite spectrum is characterized by strong
absorption lines over the UV continuum from C, N, O, Al, Si, and Fe in
various ionization stages. While some of these lines are interstellar, 
several among the strongest absorptions are identified with stellar photospheric lines.
Most of the photospheric and interstellar features are stronger in the $K$-luminous 
composite spectrum than in LBGs at $z\sim 3$. 
This suggests higher metallicity and possibly also larger interstellar velocity 
dispersion caused by macroscopic motions. The absorption lines and the slope of 
the UV continuum is well matched by the spectrum of the nearby luminous
infrared galaxy NGC~6090, which is in the process of merging. 
A metallicity higher than solar is suggested by comparing 
the pure photospheric lines (SiIII, CIII, FeV) with starburst models. The evidence of high
metallicity, together with the high masses, high star-formation rates, and possibly 
strong clustering, well qualify these galaxies as progenitors of local massive ellipticals.
\end{abstract}

\keywords{
galaxies: evolution ---
formation ---
starbursts ---
high-redshift 
}

\section{Introduction}

Despite that as much as 75\% of the stellar mass in the universe is in early-type galaxies 
(e.g. Fukugita et al. 1998), the epoch at which these massive galaxies have assembled and
acquired their high metallicity is still unknown. Although the
progenitors of these massive galaxies should be seen at high redshifts, connecting 
them with their high redshift counterparts remains a difficult task. 
The currently best known high-$z$ galaxy population is that of Lyman Break Galaxies (LBGs,
Steidel et al. 1999).
Like in local starbursts, the strongest features in the rest-frame UV spectrum of LBGs are
interstellar and photospheric absorption lines of C, N, O, Si, and Fe,
typical of hot massive stars (e.g. Shapley et al. 2003, S03 hereafter). They have
sub-solar metallicity (Pettini et al. 2001), 
very young stellar population ($<$0.2 Gyr), low masses 
($\sim10^{10}M_\odot$). Moreover, it appears that the spatial clustering of $z=3$ LBGs is too low
for them to be direct progenitors of $z\sim1$ Extremely Red Objects 
(EROs) and local massive galaxies (Daddi et al. 2001). 

While these UV-selected LBGs are certainly an important component of the high-$z$
galaxy population, it is still not clear how representative of the entire
population at high-$z$ they are. There might be other objects at high-$z$ 
that would not be picked up by this selection. In fact, there are evidences that
$K$-selected galaxies at $2<z<4$  are much more strongly clustered than 
UV-selected ones (Daddi et al. 2003). For instance, the $K$-selected galaxies
which are in the so-called {\em Redshift Desert} at $1.5<z_{\rm spec}<2.5$ 
(Daddi et al. 2004, D04 hereafter) appear to be an important class of objects,
with overall physical properties quite distinct from those of LBGs.
Comparing the properties of LBGs at $z=3$ from S03 with the sample investigated 
in this paper taken from D04, 
it appears that $K$-band luminous galaxies at $z\sim2$  
have higher reddening ($E(B-V)\sim0.3$--0.4 vs $\sim0.10$--0.17), redder UV slopes ($-\beta\sim0.54$  vs 
$\sim0.73$--1.09), 
higher SFRs ($\sim$100--500 M$_\odot$/yr vs $\sim25$--50 M$_\odot$/yr), older ages 
($\sim700$ Myr vs $\sim100$--300 Myr) and, 
therefore, higher masses ($\sim10^{11}M_\odot$ vs $\sim10^{10}$$M_\odot$).
These differences are likely arising as a consequence of the $K$-selection,
as opposed to the UV selection for LBGs.
Given their observational properties, it has been argued that these $K$-luminous galaxies 
may represent better candidates than LBGs as star forming progenitors of 
massive early-type galaxies (D04). 
If these objects are indeed the progenitors of local early-type galaxies one would expect them to 
harbor more metal rich stellar populations and interstellar medium, compared to LBGs.
In this Letter we revisit these $K$-luminous galaxies by analyzing their composite
spectrum and comparing it with those of various templates, 
including LBGs and starburst galaxies, in order to gather new
insight on their metal enrichment, on their nature and on their role in galaxy evolution.

\section{The spectra}

The K20 survey obtained VLT spectroscopy of a complete sample of about 
500 galaxies with $K_s<20$ (Cimatti et al. 2002).
In this Letter, we focus on the spectral properties of five K20
galaxies in the CDFS GOODS-South field
at $1.7<z<2.3$, with the highest S/N ratio among the 
nine presented in D04. Including the even redder K20 galaxies with 
no spectroscopic redshifts and $1.7\simlt z_{phot}\simlt2.3$, the nine objects 
represent $\sim20$--30\% of $z\sim2$ $K$-selected galaxies.
This may introduce a bias toward
the UV brightest objects, possibly minimizing the differences with respect to 
LBGs. The average spectrum nevertheless shows significantly different features when compared with 
LBGs as shown in the following sections.

The individual spectra were reduced to a common resolution of
25 \AA\ in the observed frame, by convolving them with a top-hat
function. This allowed to match the lowest resolution of some of the
spectra taken with grism 150I at $R=200$ with VLT+FORS2. 
The spectra were normalized to a common mean flux level within the range
$1300\ <\lambda < 2000$ \AA, de-redshifted and then coadded to produce
a representative average spectrum at $<z>\ =2.1$.  A flux normalized
version of the average spectrum was obtained by fitting the continuum
with a {\em spline} function.  The resolution of the
composite spectrum is low ($\sim$ 8 \AA\ in the rest-frame), however the strong
features seen in their average spectra can give important information
on the nature of these objects. 
In order to evaluate the S/N ratio of the final coadded spectrum
we used relatively line-free regions in the continuum normalized 
spectrum. In three such regions at 1270-1287 \AA, 1440-1470 \AA,
1876-1904 \AA\ we measure S/N$\sim$30 over the 1 \AA\ pixel sampling.
As all the spectra were boxcar smoothed to match the 8 \AA\ resolution
of the final average, such estimate may be taken as 
representative of the S/N ratio over a full 8 \AA\ resolution element. 
The true S/N ratio is likely higher considering that those regions will
actually contain faint lines, given that the coadded spectrum
is completely filled with absorption lines in the 1200-2000 \AA\ domain
investigated here.

\section{Comparison with Lyman Break Galaxies}

\begin{figure}
\plotone{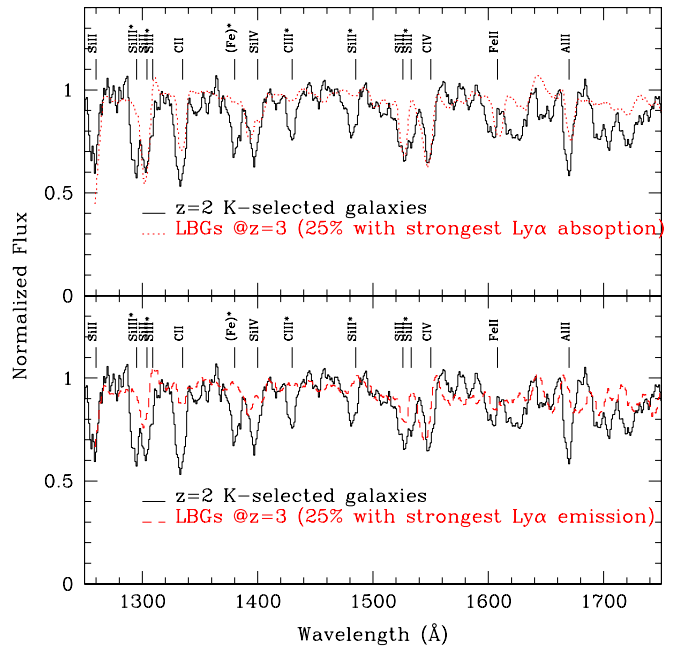}
\figcaption{Composite spectra of $K$-luminous galaxies and Lyman Break Galaxies 
(LBGs). Top and bottom panels show the LBGs with Ly$\alpha$ in absorption and emission,
respectively \citep{sha03}. 
The spectra are continuum normalized and major absorption lines 
are identified. Features marked with an asterisk are photospheric.
The resolution of the LBG spectra were degraded to that of our composite.
\label{f1}}
\end{figure}

The LBG class as a whole has been recently analyzed by S03 
using the coadded spectra of almost 1000 low S/N spectra of LBGs at z$\sim3$,
divided in 4 subsamples according to the EW of Ly$\alpha$.
In order to bracket the range of LBGs properties, 
we show in Fig.~\ref{f1} the comparison with the S03 LBGs templates having the 
lowest Ly$\alpha$ EW (i.e. Ly$\alpha$ in absorption, Group 1 in S03) and the largest
Ly$\alpha$ EW (i.e. Ly$\alpha$ in emission, Group 4 in S03). 
This latter comparison is perhaps the most appropriate, as none of the 
$K$-luminous galaxies at $z\sim 2$ have Ly$\alpha$ in emission, and also because
the LBGs with no Ly$\alpha$ emission have the reddest UV slopes (S03), although
not as red as those of the $K$-selected galaxies.

The spectrum of both high-$z$ galaxy classes is dominated by
absorption lines. The Ly$\alpha$ absorption LBG spectrum and that of the
$K$-luminous galaxies show a rather good match of the SiII~1526 and CIV~1550
features.  These are the two strongest features in the LBGs
composite spectum.  As noted by S03, the interstellar
rather than the wind component dominates the Si~IV doublet at 1400 \AA\ in the
LBGs, while the C~IV feature exhibits both the blueshifted broad
absorption and redshifted emission associated with the stellar winds,
in addition to a strong, narrower, interstellar absorption
component. The lack of resolution in our composite spectrum
smooths out the wind signatures, however, at the same resolution, 
the overall shape of the CIV feature is similar in both spectra, whereas
the SiIV interstellar line is stronger in the $K$-luminous galaxies.
Other interstellar lines such as CII~1335 and AlII~1671 are also appreciably
stronger in the
$K$-luminous galaxies than in LBGs. All absorption features are 
much weaker in the the other LBGs composites having stronger Ly$\alpha$ emission 
(Fig.~\ref{f1}, bottom panel) than in the $K$-luminous galaxies. 

The most striking differences between the LBG composite
spectra and the $K$-luminous composite are the absorption lines at $\sim$1295 \AA, $\sim$1380 \AA, 
$\sim$1430 \AA, $\sim$1485 \AA\ which are remarkably strong in the latter and 
almost undetected or very weak in the former. 
We have identified these features as the photospheric absorptions of SiIII~1296, CIII~1428, and
SiII~1485 (de Mello et al. 2000). Close to these lines there are also photospheric lines of FeII, FeV
and NIV] which could be blended with them due to the low resolution. 
The $\sim$1380 \AA\ feature is probably a blend of several Fe 
photospheric lines in the region 1360--1390 \AA\ (Leitherer et al. 2001).
In addition, the weak feature seen on the red side of SiII~1526 is
likely due to the photospheric lines SiII and FeIV at $\sim$1533 \AA, while
red broad edge of SiII~1304 seems to be due to the presence of the photospheric SiII~1309.
We interpret the strong photospheric lines in our spectrum as a
clear indication of a significantly higher metallicity in the stellar
component of the $K$-luminous galaxies at $z\sim 2$ compared to the LBGs at $z\sim 3$ 
which have sub-solar metallicity. Besides to a higher metallicity, the stronger
interstellar lines can also be
due to a higher interstellar velocity dispersion caused by large-scale
inhomogeneities and macroscopic motions in the interstellar medium,
as noted in other high-$z$ galaxies such as MS1512-cB58 (e.g. de Mello et al. 2000).

\section{Comparison with Local Starbursts}

\begin{figure}
\plotone{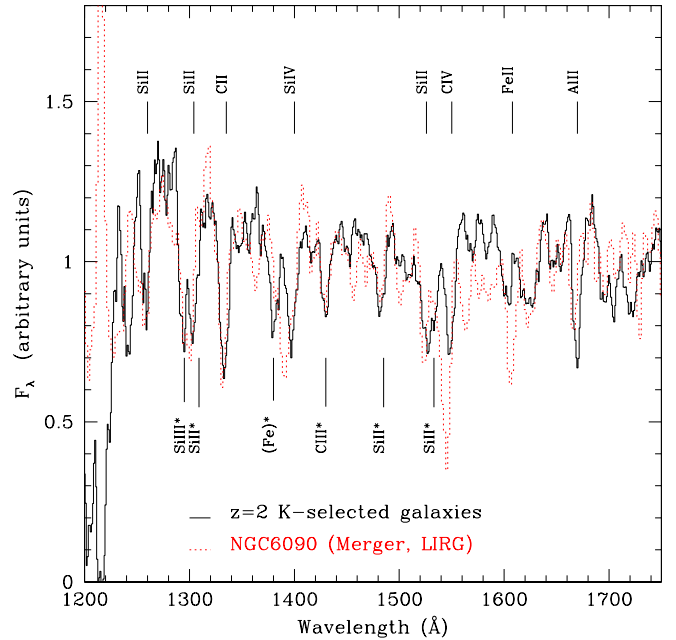}
\figcaption{The composite spectrum of $K$-luminous galaxies is compared to
NGC~6090 (Gonzalez Delgado et al. 1998). 
The spectra are overplotted with only a common scaling to the same
average flux level in 1400\AA\ to 1600\AA, with no alteration of their
intrinsic continuum shape. Major absorption lines are identified. 
The strong emission line in NGC6090 is Ly$\alpha$.
\label{f2}}
\end{figure}

The detection of several strong stellar lines in the average spectrum of 
$K$-luminous galaxies at $z\sim2$ unambiguously indicates high metallicity 
in these galaxies, to an extent that has no counterpart among other known
high redshift galaxies.
Therefore, we have checked whether a similar spectrum is exhibited by
any of the most well studied local starbursts 
galaxies.
We found that the local starbursts, NGC~1705-1, NGC~1741, NGC~4214, have very 
different spectral features than those of our $K$-selected ones.
A good match is only obtained with the spectrum of NGC~6090, both for the absorption 
lines as well as for the continuum (Fig. \ref{f2}). Note in particular that NGC~6090 
has an almost identical strength of the $\sim1295$ \AA, $\sim1430$ \AA\ and $\sim1485$ \AA\ 
absorption systems and a marginally weaker $\sim1380$ \AA\ absorption. 
These are the undetected or very weak lines in LBGs, which were identified
as photospheric lines in the previous Section.
The interstellar lines of NGC~6090 appear on average systematically blueshifted
by 1--2 \AA\ with respect to those of the $K$-luminous galaxies, unlike the stellar ones,
suggesting a weaker effect of winds. This difference is not apparent toward LBGs.
Although no strong conclusion can be derived from a single object,
this match is intriguing because the overall properties of NGC~6090 are
similar to the $K$-selected galaxies (see D04).
NGC~6090 is a well known interacting system which is in the process of merging.
It is a luminous infrared galaxy (LIRG) with log $L_{\rm IR}=11.51$ (Scoville et al. 2000) 
pointing to the presence of a strong burst of star-formation. It has a large 
number of luminous clusters triggered by the galaxy-galaxy interaction. 
The other local starbursts, on the other hand, are dwarf galaxies having likely their first
major burst of star-formation. For instance, NGC~1705-1 is a super-star-cluster (SSCs)
in a dwarf galaxy which experienced a strong burst of star-formation $\sim$10 Myr ago
(de Mello et al. 2000); NGC~1741 is an interacting system with SSCs of
a few Myr to $\sim$ 100 Myr, and masses between 10$^{4}$ and 10$^{6}$ M$_{\odot}$
(Johnson et al. 1999); NGC~4214 is a Magellanic irregular with SSCs 
of ages 1-3 Myr old (Maiz-Apellaniz et al. 1998). Like NGC~6090
the $K$-luminous galaxies are massive galaxies with spatially extended starbursts.

\section{Comparison with Starburst Models}

We have also checked whether the composite spectrum of the $K$-luminous galaxies can 
be reproduced by the Starburst99 stellar evolutionary synthesis models
(Leitherer et al. 1999; hereafter SB99). 
We first investigated the contraints given by the strength of the
CIV~1550 absorption system. In the models with instantaneous star formation 
this feature rapidly disappears after the cessation of star formation (see Fig.~5 in 
de Mello et al. 2000). Therefore, its detection in the composite spectrum 
of the $z\sim 2$ galaxies demonstrates that they contain a population of young stars. 
This supports the notion that the
red slopes of the UV spectra of the K20 star-forming galaxies at
$z\sim 2$ are due to reddening and not to ageing of their stellar
populations, confirming that these galaxies are active starbursts and thus
relevant contributors to the star formation density at $z\sim2$,
as discussed in D04.
At the same time, we do not detect the strong P-Cygni shape of CIV~1550, typical of massive, short-lived O
stars. As a consequence, a single burst which occurred within $\simlt3$ Myr can be ruled out. 
Instead, models with continuous star formation 
reproduce well the strength of CIV~1550 at pratically any older age. 

\begin{figure}
\plotone{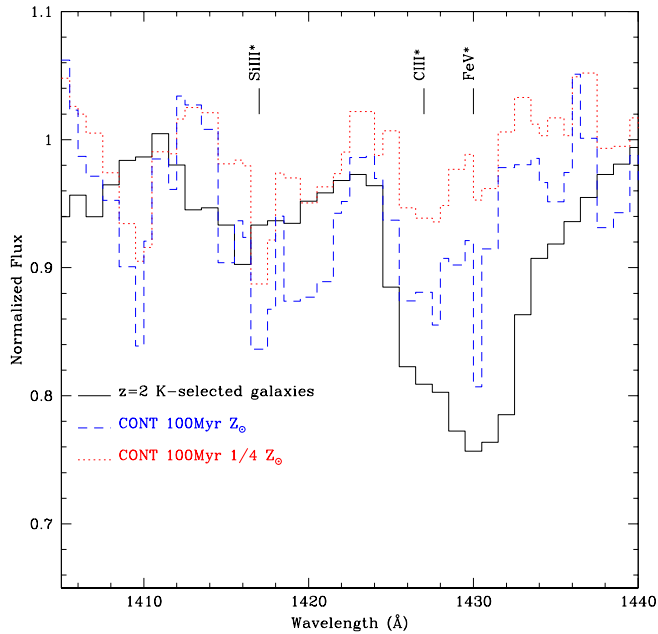}
\figcaption{Zoom of the 1430 \AA\ index region. 
We show a comparison to SB99 models: dashed lines are for solar
metallicity, dotted lines are for 0.25 Z$_{\odot}$. Spectra are
continuum normalized.  SB99 spectra are shown with their 1 \AA\ resolution.
If degraded to 8 \AA\ resolution also the SiIII~1417 absorption appears
weaker than in the $K$-luminous $z\sim2$ galaxies.
\label{f3}}
\end{figure}

Secondly, we used SB99 to obtain constraints on the overall metallicity of the 
$K$-luminous galaxies. We used the equivalent width of the absorption features 
in the wavelength range 1415-1435 \AA\ as a ``metallicity index" which 
does not strongly depend on age and IMF (Leitherer et al. 2001). 
Fig.~\ref{f3} shows a zoom of this region, 
which covers the photospheric features SiIII~1417, CIII~1427 and
FeV~1430.  As it can be seen in Fig. \ref{f3}, the three lines are
detected, but due to the low resolution CIII and FeV are blended. We
attempted to deblend the two lines by fitting the feature with two
Gaussian profiles and found that the central wavelengths of two
Gaussians do coincide with the CIII and FeV wavelengths. We
measured the equivalent width of the index to be EW$=2.3\pm0.4$ \AA. 
The same absorption system in the SB99 models has EW=0.5 \AA\ and EW=1.2 \AA\ 
for metallicities 0.25Z$_{\odot}$ and Z$_{\odot}$, respectively. 
Therefore, the $K$-luminous galaxies have metallicity index higher than the SB99 0.25Z$_{\odot}$ 
metallicity
at the 4-5$\sigma$ level and higher than the SB99 Z$_{\odot}$ metallicity at the 
$\sim$3$\sigma$ level. The 0.4 \AA\ error we derive for the metallicity index 
may be slightly underestimated because of mismatches in the continuum normalization. However, the observed 
EW of the 1415-1435 \AA\ feature would be even higher if considering also the wings of the lines 
that, because of the low resolution,
expand outside the 1415-1435 \AA\ region (EW=2.8 \AA\ in the region 1412-1440). 
Recently, Steidel et al. (2004) also used the 1425 \AA\ index and SB99 
to estimate the metallicity of UV selected galaxies at
1.4$<$z$<$2.5 and found that they have metallicity near solar.  The
fact that our sources have even stronger absorptions in this range
suggests that their metallicity is generally higher than that of Steidel et al. 
UV-selected galaxies at $z\sim 2$. This is quite plausible given that the
$K$-selected galaxies are likely to be on average more massive than the
UV-selected ones. Note, however, that the two classes of objects are not
entirely distinct, as $\sim $9\% of the UV-selected galaxies at $z\sim2$ 
are brighter than $K=20$
(Steidel et al. 2004).

Finally, we emphasize that besides the 1425 \AA\ index, other evidences
exist that point towards a high metallicity. As we showed before, there are several 
photospheric and interstellar lines that are much stronger in the 
composite spectrum than in the Z$_\odot$ SB99 models and than in the LBGs composite spectrum suggesting
that $K$-luminous galaxies have high metallicity. Similarly to Savaglio et al.
(2004), 
strong Fe~II and Mn~II lines are detected in 
the near-UV (2200-2700 \AA) region (not shown here), 
consistently with high metallicity.

\section{Discussion}

The most striking feature of the near-UV spectrum of $K$-band luminous galaxies 
at $z\sim 2$ is the presence of strong stellar absorption features, with
strengths not previously reported at high redshifts that suggest a high metallicity
for this class of high-$z$ galaxies.
Stellar photospheric lines are important diagnostic of the stellar content
of galaxies and their detection also demonstrate an UV continuum without a 
dominant AGN contribution. 
These stellar features are typical of B-type stars (de Mello et al. 2000), 
which are dominating the UV flux in the explored spectral region, rather than 
short-lived very massive O stars. This can be well expected given that the strong 
Balmer breaks detected in the near-IR imply high mass contents and continuous SFR for 
$\sim 700$ Myr (see D04). 
The fact that our composite spectrum was generated with a small number of galaxies
may affect these results, as a few peculiar objects could be present.
Coaddition of a larger number of spectra
is desirable for the future in order to smooth out the contribution of individual 
objects. 

As these $K$-selected starbursts are found to have redder UV slopes
and higher reddening than UV-selected galaxies at $1.5<z<3.5$ (the UV
selection requires blue and flat slopes), it seems reasonable to find
that they have also a higher metal content. It is not clear whether these
galaxies are evolutionary descendants of $z\simgt3$ LBGs. The
higher metallicity and dust content could be a consequence of the
continuous star formation which occurred between the two epochs. However, 
if the D04 tentative estimate of a correlation length of $r_0>7$ \h1 Mpc for the
$K$-selected galaxies at $z\sim 2$ is confirmed, they seem to be more clustered
than the LBGs at z$\sim 3$. This suggests a different evolutionary path
for the two populations. Sawicki \& Yee (1998) suggested that LBGs are 
either the progenitors of present-day sub-L* galaxies, or may form 
luminous galaxies through mergers as they evolve. However, one would
expect a stronger clustering in the latter case (Daddi et al. 2001).

The signatures of their high metallicity, the confirmation of very
high SFRs, together with the evidences for high masses and possibly
strong clustering reinforce the suggestion that these
$K$-luminous objects at $z\sim 2$ well qualify as progenitors of local
early-type galaxies, caught while still in the act of actively forming
stars. Indeed, with their masses being estimated above
10$^{11}$M$_\odot$ they would most likely become early-type galaxies 
at $z=0$.

\acknowledgments

We thank C. Steidel for providing the composite 
spectra of LBGs, S. Savaglio, D. Thomas, and R. Gonzalez-Delgado and the anonymous
referee for useful suggestions. Support for this work was provided through NASA 
grant GO09481.1 from STScI which is operated by AURA inc., under
contract NAS5-26555.

\end{document}